\documentclass[iop]{emulateapj}
\usepackage{epstopdf}
\usepackage{array}
\usepackage{multirow}

\shorttitle{HD~80606~b's Transient Heating}
\shortauthors{de Wit et al.}

\begin{document}

\title{Direct Measure of Radiative and Dynamical Properties of an Exoplanet Atmosphere} 

\author
{Julien de Wit\altaffilmark{1}, Nikole K. Lewis\altaffilmark{2}, Jonathan Langton\altaffilmark{3}, Gregory Laughlin\altaffilmark{4},\\ Drake Deming\altaffilmark{5}, Konstantin Batygin\altaffilmark{6},  Jonathan J. Fortney\altaffilmark{4}}

\altaffiltext{1}{Department of Earth, Atmospheric and Planetary Sciences, MIT, 77 Massachusetts Avenue, Cambridge, MA 02139, USA}
\altaffiltext{2}{Space Telescope Science Institute, 3700 San Martin Drive, Baltimore, MD 21218, USA}
\altaffiltext{3}{Department of Physics, Principia College, Elsah, IL 62028, USA}
\altaffiltext{4}{Department of Astronomy and Astrophysics, University of California, Santa Cruz, CA 95064, USA}
\altaffiltext{5}{Department of Astronomy, University of Maryland at College Park, College Park, MD 20742, USA}\altaffiltext{6}{Division of Geological and Planetary Sciences, California Institute of Technology, Pasadena, CA 91125, USA}

\begin{abstract}
Two decades after the discovery of 51~Peg~b, the formation processes and atmospheres of short-period gas giants remain poorly understood.  Observations of eccentric systems provide key insights on those topics as they can illuminate how a planet's atmosphere responds to changes in incident flux.  We report here the analysis of multi-day multi-channel photometry of the eccentric ($e\sim0.93$) hot Jupiter HD~80606~b obtained with the \textit{Spitzer Space Telescope}. The planet's extreme eccentricity combined with the long coverage and exquisite precision of new periastron-passage observations allow us to break the degeneracy between the radiative and dynamical timescales of HD~80606~b's atmosphere and constrain its global thermal response. Our analysis reveals that the atmospheric layers probed heat rapidly ($\sim4$-hr radiative timescale) from $<500$K to 1400K as they absorb $\sim20\%$ of the incoming stellar flux during the periastron passage, while the planet's rotation period is $93^{+85}_{-35}$ hours, which exceeds the predicted pseudo-synchronous period (40 hours).

\end{abstract}


\section{Introduction}

The discovery of 51~Peg~b \citep{Mayor1995}, the first extrasolar planet orbiting a sun-like star, was an epoch-making event. The now-famous 51 Peg~b, which has $P=4.23\,{\rm d}$ and $M\sin(i)=0.47 M_{\rm Jup}$, rapidly became the prototype {\it hot Jupiter}---gas giants on short-period orbits. In the past two decades, a substantial number of such planets have been observed and studied along their orbits \citep{Charbonneau2002,Charbonneau2005,Deming2005,Knutson2007}. The existence of hot Jupiters challenged standard planetary system formation and evolution theories developed in the context of our own solar system.

The mechanism of Kozai cycles with tidal friction, or KCTF \citep{Eggleton2001,Wu2003,Fabrycky2007}, is an attractive process to explain the current orbits of many of the known hot Jupiters. KCTF occur when a distant, sufficiently inclined stellar binary or planetary companion forces a planet on an initially long-period low-eccentricity ($e<0.1$) orbit to experience periodic episodes of very high eccentricity ($e>0.5$), a process known as Kozai cycling \citep{Kozai1962,Lidov1962}. During a planet's closest approach to its host star, periastron passage, tidal dissipation converts orbital energy into heat, thereby decreasing the planet's semi-major axis. As the planet's orbital period decreases, the Kozai oscillations are damped by general relativistic precession and other effects of comparable magnitude, marooning the planet on an inclined, gradually circularizing orbit. When orbital circularization is complete, the end product is a hot Jupiter in an orbit that is misaligned with the equator of the parent star.

\begin{figure*}[!ht]
  \vspace{-0.0cm}\centering\includegraphics[trim=0mm 0mm 0mm 00mm, clip=true,width=!,height=12cm]{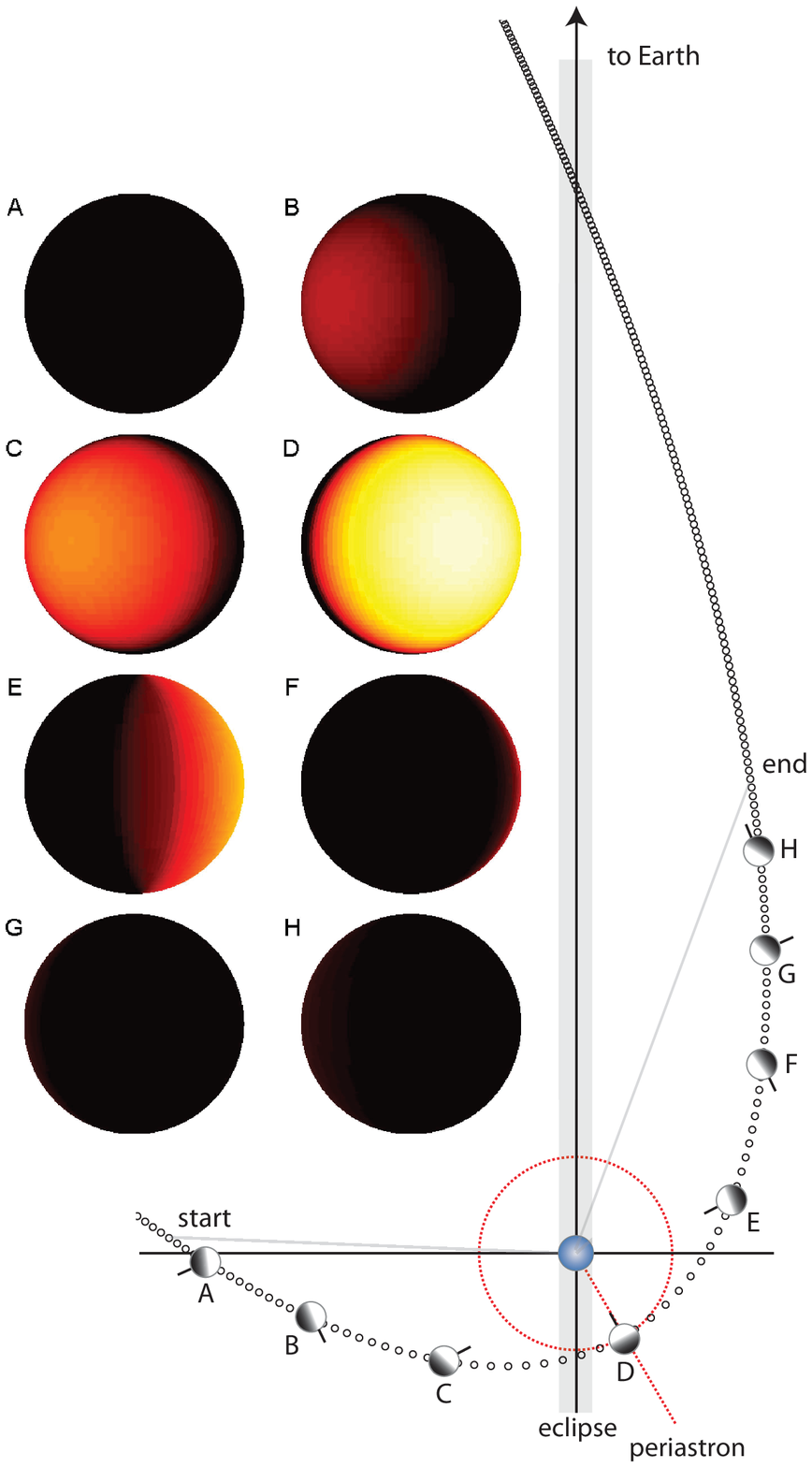}
  \vspace{-0.0cm}
  \caption{The HD~80606 system.  Dots indicate orbital positions of the planet at one-hour intervals relative to the periastron passage at HJD 2455204.91. The phase coverage of the 80-hour 4.5-$\mu$m observations is indicated by the grey lines labeled ``start'' and ``end''. The star is drawn to scale relative to the orbit.  The temperature distribution of the planet (as seen from an observer on Earth, looking down the y-axis) at times A--H is shown as the series of inset diagrams.  The brightness temperature scale runs from 500 K (black) $\rightarrow$ 1500 K (white). The \citet{Hut1981} pseudo-synchronous rotation is indicated by the black lines at the planet surface.}
  \label{fig:geomFig}

\end{figure*}

The HD~80606 system \citep{Naef2001,Laughlin2009,Moutou2009}, with its V=8.93 G5V primary and $M$=3.94$\,M_{\rm Jup}$ gas giant planet, appears to be an excellent example of a planet embarking on the final circularizing phase of the KCTF process. The external solar-type binary perturbing star, HD~80607, is plainly visible at a projected separation of $\sim$1000 {\sc AU}. The planet HD~80606~b, with a $P$=$111.43637$d orbit, has an extremely high orbital eccentricity ($e=0.93366$) and a $\lambda=42^{\circ}\pm8^{\circ}$ sky-projected misalignment between the orbital and stellar spin angular momentum vectors \citep{Moutou2009,Winn2009}, and experiences strong, episodic radiative and tidal forcing as a result of periastron encounters that sweep the planet a mere 6$R_{\odot}$ from the stellar surface (the geometry of the system is indicated in Figure\,\ref{fig:geomFig}).

Weather variations in the atmosphere of HD~80606\,b are naturally expected to be extreme.  The flux received by the planet intensifies by a factor $f=((1+e)/(1-e))^{2}=850$ between apoastron and periastron, with the major fraction of this increase coming during the 24-hour period prior to the periastron passage. The large, impulsive deposition of energy into the planetary atmosphere (equivalent to $\sim$1000x the energy delivered to Jupiter by the Shoemaker-Levy impacts) drives a global temperature increase that we have observed with \textit{Spitzer Space Telescope}'s IRAC detector \citep{Werner2004}.

\section{Observations and Data Reduction}

Using the 4.5-$\mu$m channel of the IRAC detector in subarray ($32\times 32$ pixel) mode, we obtained a series of 676,000 photometric measurements of HD~80606 in two successive blocks of 50 and 30 hrs starting on UT Jan 06 2010 and ending on Jan 09 2010, with a 0.4 sec exposure time. The first block started 34 hours before the periastron passage at HJD 2455204.91. The second block started after a three-hour interruption for data transfer following the acquisition of the first block. In addition, we have re-analyzed a separate (shorter) set of photometric measurements obtained in 2007 using \textit{Spitzer}'s 8-$\mu$m channel \citep{Laughlin2009}.

\subsection{\textit{Spitzer} 4.5-$\mu$m Photometry}
We calculate the time at mid-exposure for each image from time stamps stored in the FITS header. For each FITS file, we determine the background level of each of the 64 subarray images from the pixels located outside a 10 pixel radius, excluding any flagged hot,bright or peripheral pixels. Using a histogram of the measured background pixel values we discard the 3$\sigma$ outliers from the background pixels. We estimate the background value based on the median of the undiscarded background pixels. We then subtract this background value from each image. From the background subtracted image, we determine the stellar centroid based on a flux-weighted centroid method using a circular aperture with a radius of 4 pixels \citep{Knutson2008,Lewis2013}. We estimate the stellar flux from each background subtracted image using a circular aperture centered on the estimated position of the star. We optimize our choice of aperture individually for each AOR looking both at the white and red noise contributions for time-varying and fixed apertures \citep{Knutson2012, Lewis2013}. We find that for all the AORs time-varying apertures equal to the square root of the noise pixel parameter $\tilde{\beta}$ with a constant offset minimize the scatter and the relative amount of red noise in the final time series and have an average aperture of 1.8 pixels.
   
Finally, we remove outliers from our final time series sets using 4 $\sigma$ moving filters with widths of 64, 1000, 2000, and 4000 on the photometry and the centroid position. The use of multiple large filters allow us to remove pointing excursions with longer ($\sim$ 1 minute) durations, such as those due to micrometeorite impacts on the spacecraft. We find that these pointing excursions are relatively rare, and these large filters account for less than 15\% of the total outliers. Overall, we discard 0.7\% of the images.

\subsection{\textit{Spitzer} 8.0-$\mu$m Photometry}

The same procedure is applied to reduce the 8.0-$\mu$m photometry, although the observations are obtained in fullarray mode. We find that for all the 8-$\mu$m AORs a fixed aperture equal to 2.7 pixels minimize the scatter and the relative amount of red noise in the final time series. Overall, we discard 4.7\% of the images.

\section{Photometry Analysis}

\begin{figure*}[!ht]
  \vspace{-0.0cm}\centering\includegraphics[trim=0mm 0mm 0mm 00mm, clip=true,width=17cm,height=!]{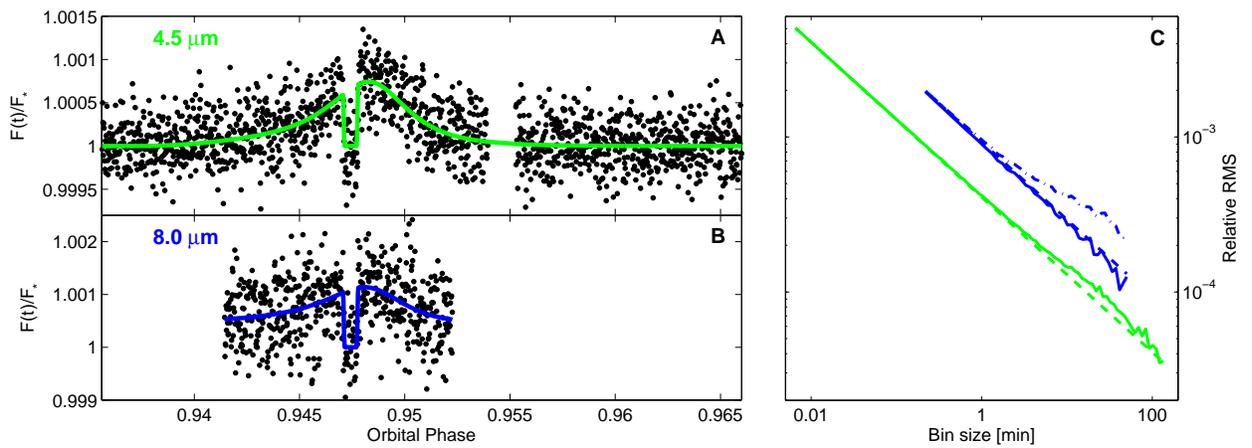}
  \vspace{-0.0cm}
  \caption{\textit{Left}: HD~80606's photometry at 4.5 \textbf{(A)} and 8.0~$\mu$m \textbf{(B)}. The photometry (black dots) is shown with the instrumental effects removed and binned into three minute interval. We overplot the best-fit model light curve as plain lines (in green and blue for the 4.5 and 8.0-$\mu$m channels, respectively). \textit{Right}: Standard deviation of the best-fit residuals vs. bin size for the 4.5 and 8.0-$\mu$m HD 80606 observations \textbf{(C)}. The standard deviations of residuals vs. bin size are shown as plain lines for each channel together with the expected behavior of Gaussian noise with binning (dashed lines). The dashed-dotted blue line shows the same quantity for the best-fit light-curve obtained at 8~$\mu$m while not accounting for intrapixel sensitivity variation.}
  \label{fig:best_fits}
  
\end{figure*}

The photometric reduction process for these data is complicated by a number of well-documented observational artifacts that combine  to generate spurious $\sim$1\%-level fluctuations in the raw photometry. We correct for these instrumental systematics using standard techniques described below allowing us to reach photometric precisions of 55 and 100 ppm per 1-hr bin at 4.5 and 8~$\mu$m, respectively. The resulting 0.1\% flux variation of the system during the course of the observations is shown in Figure\,\ref{fig:best_fits}.

We analyze HD~80606's photometry using a method previously introduced in the literature \citep{deWit2012,deWit2015}. The method is implemented as an adaptive Markov Chain Monte Carlo (MCMC) \citep{Gregory2005,Ford2006} algorithm, allowing us to stochastically sample the posterior probability distribution for our chosen model, providing an estimate of the best-fit model parameters, their uncertainties, and any covariances between parameters.  We update the method to include the transient heating of the planet's atmosphere due to its eccentric orbit.

\subsection{Nominal Model}
\label{sec:model}

We use a simple energy-conserving model of the planet's global atmospheric response to estimate the atmospheric properties of HD~80606~b by fitting the time series data. The model assumes that the photometry is unaffected by any advective hydrodynamical response, and that the planet has a uniform global temperature, $T_0$, 50 hrs prior to its periastron passage, which is consistent with predictions from global circulation models for exoplanets on high eccentricity orbits \citep{Langton2008, Kataria2013, Lewis2014}. For both the 4.5-$\mu$m and the 8-$\mu$m channels, at each point on a planetary latitude ($\theta$) longitude ($\phi$) grid, and at each time stamp, we compute the rate of temperature change $\dot{T}$ using
\begin{equation}
\dot{T} = \gamma_T (T_{\rm eq}^4 - T^4),
\end{equation}
where $\gamma_T$ is a parameter related to the radiative response rate and $T_{\rm eq}(t, \phi, \theta) $ is the temporally- and spatially-varying equilibrium temperature of the atmosphere.  Specifically, we take
\begin{equation}
T_{\rm{eq}}^4=f \left(\frac{L_*}{4 \pi \sigma r^2}\right) \cos \alpha+T_0^4.
\end{equation}

In the above equation, the instantaneous star-planet distance, $r$, is  $r=a(1-e\cos E)$, where $a$ is the semi-major axis and $E$ is the eccentric anomaly, which is related to the mean anomaly, ${\cal M}$, through Kepler's equation, ${\cal M}=E-e\sin E$.
The absorptivity, $f$, is the fraction of the incoming stellar flux absorbed at the photospheric layer, $\sigma$ is the Stefan-Boltzmann constant and $\alpha=\cos(\theta) \cos(\phi-2\pi t/P_{\rm rot})$ is the zenith angle of the star.  Following \citet{Iro2005,Seager2005}, we derive the radiative timescale near the photosphere as
\begin{equation}
\tau_{\rm rad}(T)=\frac{1}{\gamma_T T^3}.
\end{equation}
In order to report a single radiative timescale for the entire planet, we compute $\tau_{\rm rad}$ from $\gamma_T$ after adopting a fiducial value, ${\bar T}=1200$ K---which correspond to an average of the temperatures reached locally within the irradiated part of HD~80606~b's photosphere during the periastron passage.

Our model thus incorporates four measurable parameters for each channel, each with clear physical meaning: (1) $\gamma_T$ that relates to the radiative timescale $\tau_{\rm rad}$, which depends on the atmospheric depth of the infrared photospheres; (2) the fraction $f$ of incoming flux absorbed at the photospheric layer, (3) the planetary rotation period $P_{\rm rot}$, which determines the substellar longitude at any given time; and (4) the temperature out-of-irradiation, $T_0$, which constrains the contribution of tidal heating to the atmosphere's energy budget. As our  data contain only phase-curve and secondary-eclipse measurements, we fix the values of the planet's orbital configuration to the one estimated in \cite{Hebrard2010}.

\subsection{Pixelation Effect and Intrapixel Sensitivity Variations}

The change in position of the target's point-spread function (PSF) over a detector with non-uniform intrapixel sensitivity leads to apparent flux variations that are strongly correlated with the PSF position. Studies have demonstrated that an empirical ``pixel mapping'' method produces optimal results for bright stars and longer time series observations \citep{Ballard2010,Stevenson2012,Lewis2013}. Here, we use the same implementation of the pixel-mapping method as in \citet{Lewis2013}. 

We find that accounting for intrapixel sensitivity also improve significantly the correction of observations at 8~$\mu$m (Figure\,\ref{fig:best_fits}.C), as previously pointed out by \citet{Stevenson2010}. In particular, we find that accounting for such systematics decreases the red noise contribution by 66\% (i.e., more than a factor two smaller).

\subsection{Detector Ramp}

IRAC observations often display an exponential increase in flux at the beginning of a new observation.  Similarly to \citet{Lewis2013}, we trim the first hour at the start of each 4.5-$\mu$m observation and subsequent downlinks which reduces the complexity of our fits by removing the exponential trend with minimal loss of information on the eclipse shape and time.  Similarly to \citet{Agol2010}, we correct for the 8-$\mu$m ramp using a double exponential.

\subsection{Time-Correlated Noise and Uncertainty Estimates}

We find that the standard deviation of our best-fit residuals is a factor of 1.09 and 1.19 higher than the predicted photon noise limit at 4.5 and 8~$\mu$m, respectively. We expect that this noise is most likely instrumental in nature, and account for it in our error estimates using a scaling factor $\beta_{red}$ for the measurement errobars \citep{Gillon2010a}. We find that the average $\beta_{red}$ in the 4.5 and 8~$\mu$m channels are 1.3 and 1.15 (Figure\,\ref{fig:best_fits}.C).

\section{Results \& Discussion}

\begin{figure*}[!ht]
  \vspace{-0.0cm}\centering\includegraphics[trim=0mm 0mm 0mm 00mm, clip=true,width=!,height=9cm]{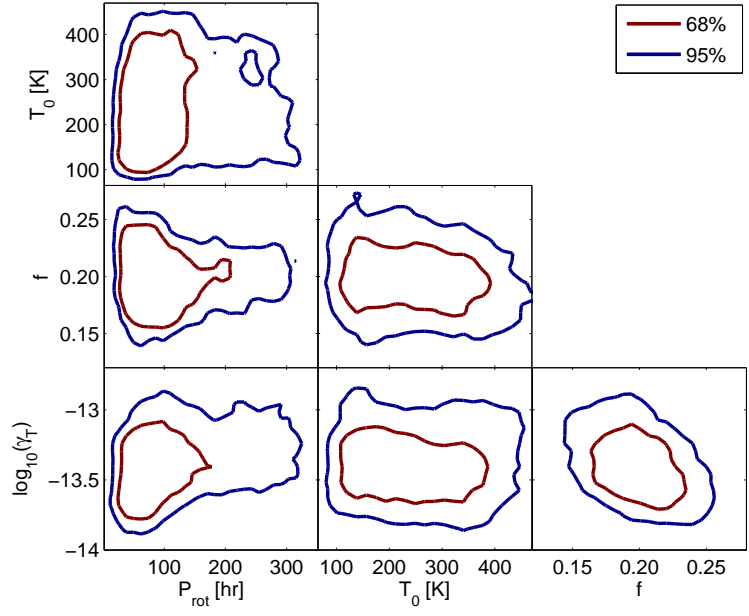}
  \vspace{-0.0cm}
  \caption{Posterior distribution of the atmospheric model parameters (Section\,\ref{sec:model}). The contours encompasses 68\% and 95\% of the distribution obtained from a global fit using the same model for HD\,80606\,b's 4.5 and $8.0\mu$m photospheres. The distribution shows no significant correlation between the model parameters.}
  \label{fig:2D_DDP}

\end{figure*}

Inspection of the light curve reveals that there is a rapid rise in infrared flux from the planet during the hours prior to the periastron passage. The occultation of the planet establishes a baseline stellar flux, and demonstrates that the planet is not radiating strongly enough to be detectable at the start of the 4.5-$\mu$m time series. Following periastron passage, both the  4.5-$\mu$m and the 8-$\mu$m time series display rapid flux decreases from the planet. The waning planetary flux can be attributed to a combination of photospheric cooling and the rotation of the heated atmospheric area from the line of sight. The final 30 hours of the time series show no evidence of any resurgence of flux from the planet implying that the radiative timescale of the atmospheric layer probed at 4.5~$\mu$m is significantly shorter than the planet's rotation period. Table\,\ref{tab:fit_param} gathers the key quantities constrained in the present study. We present in Figure\,\ref{fig:2D_DDP} the posterior distribution of the atmospheric model parameters that shows no sign of correlation. The plotted distribution is obtained from a fit using the same atmospheric model parameters for HD~80606~b's 4.5 and 8.0-$\mu$m photospheres as the consistency of their individual estimates suggest overlapping between both photospheres. 

We find that $P_{\rm rot}=93^{+85}_{-35}$ hours for HD~80606~b, which provides an observational insight into the relation between spin and orbit rotation rates of eccentric planets. In contrast to planets on circular orbits, eccentric planets experience a widely-varying orbital velocity that prevents the planets from achieving spin-orbit synchronization. It is expected that tidal forces drive eccentric planets into a state of {\it pseudo-synchronous} rotation in which the spin frequency is of order the instantaneous circular orbital frequency near periastron. The model that is generally adopted for this process is that of \citet{Hut1981}, which draws on the equilibrium tidal theory of \citet{Darwin1908}, (and which was qualitatively invoked by \citet{Peale1965} to explain Mercury's asynchronous spin).
For HD~80606~b, this period is $P\,$= 39.9 hr, which is in mild tension with our measurement. Although there exist alternative theories to \citet{Hut1981}, these suggest rotation periods smaller than \citet{Hut1981}. For example, \citet{Ivanov2007} argue that the spin frequency of a pseudo-synchronous satellite should be 1.55 times the circular orbital frequency at periastron, yielding $P_{\rm rot}=28$h.  The measured rotation period of HD~80606~b could imply either a recent initiation of the synchronization process or that its efficiency is lower than predicted. Simultaneously, we note that our derivation of the rotation rate from the light curve may be contaminated by additional physical effects (such as changes in the atmospheric composition) for which our model does not account. While such detailed modeling is beyond the scope of this work, it constitutes an important avenue for future analysis. 

\begin{figure*}
  \begin{center}
  	\includegraphics[trim = 00mm 20mm 00mm 00mm,clip,width=13cm,height=!]{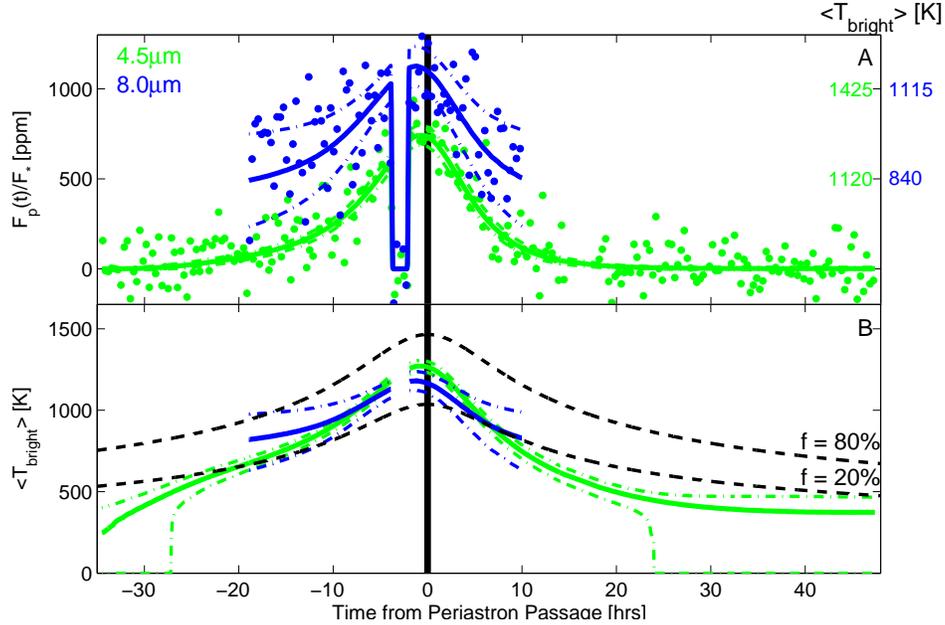}
    \vspace{-0.2cm}
    \end{center}
  \caption[HD 80606 b's transient heating observed in \textit{Spitzer}/IRAC 4.5 and 8.0-$\mu$m bands.]{HD 80606 b's transient heating observed in \textit{Spitzer}/IRAC 4.5 and 8.0-$\mu$m bands. \textbf{(A)} HD 80606 b's relative thermal fluxes in \textit{Spitzer}/IRAC 4.5 and 8.0-$\mu$m bands as a function of the time from periastron passage. Quantities shown in green and blue refer to measurements obtained at 4.5 and 8.0~$\mu$m, respectively. The photometry corrected for the instrumental effects and binned into twenty-minute intervals is shown as filled dots. The median models and their 1$\sigma$ envelope are shown as plain and dashed-dotted lines, respectively. The larger extent of the 1$\sigma$ envelope at 8.0-$\mu$m is due to a uncertainty on the measurements that is 2.1 times larger than at 4.5-$\mu$m and to a smaller time coverage combined with instrumental effects. Conversion of the relative planetary fluxes to hemisphere-averaged brightness temperature are shown on the right for each channel. \textbf{(B)} HD 80606 b's hemisphere-averaged brightness temperature as a function of the time from periastron passage derived from the planetary fluxes introduced in \textbf{(A)}. The dashed lines represent the equilibrium temperature of a planet assuming full and instantaneous energy redistribution for an absorptivity of 0.8 and 0.2 \citep{Seager2010}.} 
  \label{fig:data}
\end{figure*}

Our measurements reveal that the global temperature of the probed atmospheric layers increases rapidly from $<500$K to 1400K during the periastron passage (Figure\,\ref{fig:data}). The shape of the planetary fluxes as a function of time---including the shift of the peak flux from periastron passage and overall symmetry of the response---implies that the photospheric layers at 4.5 and 8~$\mu$m absorb $19.9\pm2.1$ and $13.6\pm4.5$\% of the incoming stellar flux and have a radiative timescales of 4.04$\pm^{2.57}_{1.55}$ and 1.28$\pm^{6.08}_{1.04}$ hours, respectively.  The hemisphere-averaged brightness temperatures (Figure\,\ref{fig:data}.B) and the parameter estimates for the 4.5 and 8~$\mu$m photospheres are consistent at the 1-sigma level suggesting significant overlap in the atmospheric pressure levels probed by each bandpass.  The atmospheric properties estimated near the 4.5 and 8.0~$\mu$m photospheres suggest that pressures near the $\sim$10-100~mbar and $\tau=1/5$ level in the optical are being probed by these bandpasses.  HD~80606~b's marginal brightness through the bulk of its orbit implies either the presence of significant cloud coverage that suppresses the planet's observable thermal emission, similarly to \citet{Demory2013}, or a marginal thermal emission. Assuming HD~80606~b's atmosphere to be cloud free, this implies that HD~80606~b's atmosphere undergoes a transient thermal inversion during the periastron passage as predicted in \citet{Lewis2014}. Furthermore, the Stefan-Boltzmann law implies a tidal luminosity $L_{\rm tid}<4\pi R_p^2\times \sigma T_{0}^4\lesssim 2\times10^{20}$ W. Assuming that the tidal heat is primarily dissipated in the deep interior and dominates other internal heat sources, this value can be used to place an upper limit on the planet's tidal quality factor $Q$, through the relation \citep{Wisdom2008}:
\begin{equation}
L = \frac{21}{2}\frac{k_2}{Q}\frac{G M^2 R^5 n}{a^6} \zeta_{L}(e)\,,
\label{Ltid}
\end{equation}
giving $Q>2.5\times10^6$. This value would indicate that HD~80606~b is much less dissipative than any of the solar system planets. For example, estimates of the tidal quality factor of Jupiter give $Q_{\rm Jup}=3\times10^{4}$, based on measurements of the secular accelerations of the Jovian satellites \citep{Lainey2009}. A comparatively low rate of tidal energy dissipation within HD~80606~b would then be consistent with it having retained a large eccentricity and a large rotation period in the face of billions of years of tidal evolution. 
\
\section{Conclusions}

Our analysis of \textit{Spitzer}'s phase curve data for HD~80606~b has provided critical insights into the planet's atmospheric and interior properties.  Data analysis techniques and our understanding of systematics for \textit{Spitzer} observations have greatly improved since the original reduction of the 8-$\mu$m data by \citet{Laughlin2009}, which provided the first estimate for radiative timescales in hot Jupiter atmosphere.  By reanalyzing the 8-$\mu$m  data and combining it with a longer baseline 4.5-$\mu$m data set, we obtained independent constraints on the radiative and dynamical timescales and absorptivity of HD~80606~b's 4.5 and 8-$\mu$m photospheres. Our new estimates of the radiative timescales are consistent with \citet{Laughlin2009} and support inefficient transport of energy from the dayside to the nightside of the planet---the observations presented here do not provide further constraints on HD~80606~b's atmospheric dynamics.  The long temporal baseline of the 4.5-$\mu$m data also allowed for constrains on the internal temperature of HD~80606~b, which together with those on its rotation period have challenged assumptions generally made on tidal dissipation in hot Jupiters. Although it is plausible that additional processes such as cloud formation play a role in shaping HD~80606~b's phase-curve, the planet's response near periastron passage will be dominated by radiative processes that are well described by our simple physical model.  A transient thermal inversion will form near the probed photospheres (above any cloud decks) that will dominate the planetary flux signal from which the absorptivity, radiative timescale, and rotation rate were derived. Therefore, only the insights derived from the $T_0$ parameter may be affected. In order to get further insights into HD~80606~b's atmospheric and formation processes, complementary observations at shorter and longer wavelengths are required to constrain the extent to which scattering and emission processes are shaping its flux near periastron.

\acknowledgments

JdW thanks the Zoweh crew for their hospitality during part of this work. This work is based on observations made with the \textit{Spitzer Space Telescope}, which is operated by the Jet Propulsion Laboratory, California Institute of Technology, under contract to NASA. Support for this work was provided by JPL/Caltech.

\bibliographystyle{Biblio/apj}

\begin{table*}[!b]
	\caption{HD 80606 b's atmospheric parameters constrained from phase curve measurements.}
	\label{tab:fit_param}
		
	\setlength{\extrarowheight}{2pt}
	\centering\footnotesize{\begin{tabular}{c|cc}
		\hline\hline
		& \textbf{4.5~$\mu$m} & \textbf{8.0~$\mu$m}\\
		
		\hline

Eclipse depth [ppm] & 651$\pm$49 & 1053$\pm$94\\
Phase-curve peak [ppm] & 738$\pm$52 & 1105$\pm$121\\
Absorptivity ($f$) & 0.199$\pm$0.021 & 0.136$\pm$0.045\\
Radiative timescale ($\tau_{\rm rad}$) [hr] & 4.04$\pm^{2.57}_{1.55}$ & 1.28$\pm^{6.08}_{1.04}$\\
3$\sigma$ upper limit on $T_0$ [K] & 493 & 1205\\
Rotation period ($P_{\rm rot}$) [hr] & \multicolumn{2}{c}{93$\pm^{85}_{35}$}\\
\\
		\end{tabular}}


\end{table*}

\end{document}